\documentclass[11pt]{article}

\usepackage{authblk}
\usepackage{geometry}
\usepackage{amsmath}
\usepackage{xcolor}
\usepackage{etoolbox}
\usepackage{amsfonts}
\usepackage{amssymb}
\usepackage{graphicx}
\usepackage{float}
\usepackage{tablefootnote}
\usepackage[]{breqn}
\providetoggle{long}
\settoggle{long}{false}

\providecommand{\keywords}[1]
{
  \small	
  \textbf{\textit{Keywords---}} #1
}

\author[1]{Nicholas T. Layden \thanks{Nicholas.Layden@dal.ca}}
\author[2]{Alan A. Coley \thanks{Alan.Coley@dal.ca}}
\author[3]{Dipanjan Dey \thanks{deydiprohan@gmail.com}}

\affil[1,2,3]{Department of Mathematics and Statistics, Dalhousie University, Halifax, Nova Scotia, Canada B3H 3J5}

\title{Invariant description of static and dynamical Brans-Dicke spherically symmetric models}
\date{}

\begin{document}
	\maketitle

\begin{abstract}
	We investigate spherically symmetric static and dynamical Brans-Dicke theory exact solutions using invariants and, in particular, the Newman Penrose formalism utilizing Cartan scalars. In the family of static, spherically symmetric Brans-Dicke solutions, there exists a three-parameter family of solutions, which have a corresponding limit to general relativity. This limit is examined through the use of Cartan invariants via the Cartan-Karlhede algorithm and is additionally supported by analysis of scalar polynomial invariants. It is determined that the appearance of horizons in these spacetimes depends primarily on one of the parameters, $n$, of the family of solutions. In particular, expansion-free surfaces appear which, for a subset of parameter values, define additional surfaces distinct from the standard surfaces (e.g., apparent horizons) identified in previous work. The ``$r=2M$" surface in static spherically symmetric Brans-Dicke solutions was previously shown to correspond to the Schwarzschild horizon in general relativity when an appropriate limit exists between the two theories. We show additionally that other geometrically defined horizons exist for these cases, and identify all solutions for which the corresponding general relativity limit is not a Schwarzschild one, yet still contains horizons. The identification of some of these other surfaces was noted in previous work and is characterized invariantly in this work. In the case of the family of dynamical Brans-Dicke solutions, we identify similar invariantly defined surfaces as in the static case and present an invariant characterization of their geometries. Through the analysis of the Cartan invariants, we determine which members of these families of solutions are locally equivalent, through the use of the Cartan-Karlhede algorithm. In addition, we identify black hole surfaces, naked singularities, and wormholes with the Cartan invariants. The aim of this work is to demonstrate the usefulness of Cartan invariants for describing properties of exact solutions like the local equivalence between apparently different solutions, and identifying special surfaces such as black hole horizons.
\end{abstract}

\section{Introduction}
The gravitational collapse of matter is an important research area in general relativity (GR). An essential
question is whether and under what initial conditions a black hole
or naked singularity (NS) forms in the gravitational collapse \cite{Dey1, Dey2}. A model system for studying this question is provided by Brans-Dicke  (BD) theory \cite{BT}, which is formally equivalent to the Einstein equations minimally coupled to a massless scalar field. The simplified set of equations obtained by imposing spherical symmetry is easier to handle.
There are a number of exact solutions known for this system, almost all of
which are either static or depend only upon the time coordinate.

The Brans-Dicke theory is a scalar-tensor theory of gravitation, whose field equations are given by:

\begin{equation}
\begin{aligned}
	G_{\mu\nu} &= \frac{8\pi}{\phi}T^{(m)}_{\mu\nu} + \frac{\omega}{\phi^2}\left(   \nabla_{\mu} \phi \nabla_{\nu} \phi - \frac12 g_{\mu\nu}\nabla^{\lambda}\phi\nabla_{\lambda}\phi \right) + \frac{1}{\phi}(\nabla_{\mu} \nabla_{\nu} \phi - g_{\mu\nu} \Box \phi) - \frac{V}{2\phi}g_{\mu\nu},\\ 
	\Box \phi &= \frac{1}{2\omega + 3}\left( 8\pi T +\phi \frac{dV}{d\phi}-2V \right),
\end{aligned}
\end{equation}

\noindent
where $T^{(m)}_{\mu\nu}$ is the stress-energy tensor for the matter content, $\phi$ is the Brans-Dicke scalar field, $T\equiv T^{\mu}_{\ \mu}$ is the trace of the matter tensor, $V(\phi)$ is the scalar field potential, which can be taken as a generalization of the cosmological constant in GR, and $\omega$ is the Brans-Dicke parameter. The d'Alembertian operator is defined as $\Box \equiv \nabla_{\mu}\nabla^{\mu}$. For the purposes of this paper, we investigate models which have $V(\phi)=0$ and $T=0$.

\subsection{Geometry}

For stationary black hole solutions, the event horizon can be identified with the Killing horizon (KH), which is a local surface. The equation often used for locating an apparent horizon (AH) is  \cite{Hochberg:1998ha}: $g^{ab}R_{,a}R_{,b} =0$, where R is the areal radius. However, it is of interest to study hypersurfaces defined independently of the choice of foliation. Black hole spacetimes with a geometrically defined quasi-local horizon, referred to as a {\it geometric horizon} (GH) were discussed in \cite{COLEY2017131}, in which the curvature tensor is algebraically special relative to the alignment classification \cite{classb}. 
A GH can be invariantly defined by the vanishing of a particular set of scalar polynomial curvature invariants (SPIs)  \cite{GH}, which are truly frame-independent. In the case of spherically symmetric black holes, the AH is a GH and is unique  \cite{GH}.

The definition of a GH can also be given in terms of Cartan invariants \cite{kramer}, which must be calculated in a certain prescribed invariant coframe \cite{COLEY2017131} within the Newman-Penrose (NP) formalism. A GH implies that $\rho = 0$, defining an expansion free surface (where $\rho$ is an NP spin coefficient). Such EF surfaces are of interest here. In spherically symmetric models, it is common to associate the spin coefficient $\rho$ with the expansion of a null geodesic congruence, though $\rho$ can take complex values, so we must be careful with our exact interpretation. Note that a KH and wormhole throat is also detected by $\rho = 0$. Indeed, SPIs have previously been used to examine the curvature structure of black hole, naked singularity, and wormhole solutions \cite{McNutt}. Conformal invariant quantities of scalar-tensor theories have also been studied \cite{Kozak,mcpage}.

The limit of BD spacetimes as the scalar field coupling constant $\omega$ tends to infinity applying a coordinate-free technique using Cartan scalars
has been investigated \cite{Paiva}. Applications to the dynamics of spherically symmetric thin shells have also been studied in the framework of BD theory using the NP formalism \cite{Letelier}. We will apply these geometric techniques here to examine the properties of families of Brans Dicke static and evolving spherically symmetric spacetimes.

\subsection{Static Models}

Static spherically symmetric solutions of the vacuum BD theory have been studied for all values of the BD parameter $\omega$  \cite{Brans62}. In GR there is a {\emph{unique}} spherical
and asymptotically flat solution of the vacuum Einstein
equations (with zero cosmological constant): the static
Schwarzschild geometry. This is generally not the case in
theories of gravity alternative to GR.
The prototypical alternative to GR is BD
gravity \cite{BT}, which adds to the metric tensor
a scalar
degree of freedom, $\phi$, which continues to garner interest.
However, Hawking  \cite{Hawk} showed that  all
vacuum, static, and asymptotically flat black holes
of BD gravity must reduce locally to the Schwarzschild black hole.
An essential feature in the proof of this relies upon the fact that $\phi$
becomes constant outside the horizon, reducing the theory to GR. BHs have constant scalar field outside their horizons. In the limit $\omega \to \infty$, the
BD theory apparently reduces to GR. However, it has been shown that as the scalar field coupling constant $\omega\to \infty$, some solutions of the BD equations for a specified matter configuration do not go over to a GR solution for the same energy-momentum tensor \cite{Paiva}. The existence of a GR limit for a particular BD solution is connected to the vanishing of the trace of the stress energy tensor for the matter component \cite{val}. Whether there exists an appropriate limit from a BD theory solution to an equivalent GR theory solution has been discussed in \cite{val}.

{\em{The general
form of the non-black hole solutions of the general static, spherically symmetric, asymptotically flat,
vacuum BD field equations are given by the Campanelli-Lousto (CL)  \cite{campanelli93} family, which describe wormhole throats and naked singularities \cite{Bhadra2005,Agnese95} (see also \cite{Bron,val}).}}

	\begin{equation}\label{bdss}
		ds^2 = -\left(1-\frac{2M}{r}\right)^{m+1}dt^2 + \left(1-\frac{2M}{r} \right)^{n-1} + r^2\left(1-\frac{2M}{r}\right)^{n}(d\theta^2 + \sin(\theta)^2d\varphi^2),
	\end{equation}
	
	\begin{equation}\label{bdss_sf}
		\phi = \phi_0 \left(1-\frac{2M}{r} \right)^{-(m+n)/2},
	\end{equation}
	
	\noindent
	where $m,n$ are arbitrary constants, which are related to the BD parameter $\omega$ by
	
	\begin{equation}
		\omega = -2\frac{m^2 + n^2 + nm + m - n}{(m+n)^2}.
	\end{equation}
	
	\noindent
	The GR limit of this metric corresponds to $m\to-n $, which implies $\omega \to -\infty$.

The degenerate case of  BHs and the GR limit are not covered 
by the class of solutions (\ref{bdss}). The Schwarzschild black hole is indeed a solution and is easy to verify \cite{val}. It was demonstrated in \cite{campanelli93} that for some particular choices of the solution parameters \cite{Bhadra2005}, the BD theory admits static and spherically symmetric black holes different from the Schwarzschild solution when the weak energy condition is not respected. These results about the nature of the solutions of BD theory have already been discussed in \cite{val}.

In more detail, to assess whether the general
geometry describes black holes, wormholes, or naked
singularities, one examines the horizons (if they exist)
and their nature. The equation often used for
locating the apparent horizons is \cite{Hochberg:1998ha}:  $g^{ab}R_{,a}R_{,b} =0$, where $R$  is the areal radius. Horizons correspond to the roots of
that equation; a single root describes a black hole horizon
while a double root describes a wormhole throat. 
In the CL parameterization, if roots exist (corresponding to a range of parameter values), they are always double
roots corresponding to wormhole throats. Otherwise (for other parameter values), instead, there is a naked singularity. Note that in all cases, the metric has a spacetime singularity at $r = 0$ (as is deduced from the Ricci scalar).

It can be seen that (for certain parameter values) as $r \to 2M$ all curvature invariants diverge and the solution exhibits a naked singularity. 
The locus of points at $r=2M$  is a timelike curvature
singularity, and so the `horizon' is shrunk to
a point as in the static metric given by others 
\cite{Agnese95,campanelli93,Bhadra2005,val}.
However, for certain parameter ranges (e.g., corresponding to negative values of $\omega$), the
curvature invariants become non-singular, whence the surface $r = 2M$ will
be an outgoing expansion-free surface and hence it will act as the event horizon and the solution exhibits black hole nature \cite{campanelli93}. 

\subsubsection{Killing Horizons}
	Since this is a static solution, we can investigate the Killing vectors of the metric, and examine the null hypersurfaces generated by them. There exists only a single timelike Killing vector, given by $\xi =\partial_t$, with the following norm:

	\begin{equation}
		\xi^a \xi_a = \left(1-\frac{2M}{r} \right)^{m+1},
	\end{equation}
	
	\noindent
	where there is a Killing horizon (KH) when the parameter $m$ satisfies $m>-1$, at $r=2M$.

\subsubsection{Wormholes}

Wormholes are objects that connect two or more different regions of spacetime, or even separate universes, whose distinct regions are bounded by a wormhole throat \cite{morris88, Visser:1995cc}. Their definition relies on the global geometry. The wormhole definition that is implicitly adopted here is that of \cite{Hochberg:1998ha}, which consists of a quasi-local definition involving only the properties of the local geometry of spacetime. Invariants have previously been used to examine the curvature structure of wormhole solutions \cite{Agnese95,McNutt}. Traversable wormholes have also been investigated \cite{Hochberg:1998ha}.

In \cite{McNutt} a spherically symmetric line element which admits either a black hole geometry or a wormhole geometry was considered and it was shown that in both cases the AH or the wormhole throat is partially characterized by the zero-set of a single curvature invariant. To distinguish between these surfaces, we determine a set of curvature invariants that fully characterize the AH and wormhole throat. The definition of a wormhole throat has an issue with foliation dependence. In the case of stationary or static wormhole solutions, the throat can be determined locally \cite{Hochberg:1998ha}. However, for dynamical wormholes there are several throat definitions that arise from imposing conditions on the expansion of ingoing or outgoing null geodesics and their derivatives; in particular,
\cite{Hochberg:1998ha} described the wormhole throat as the minimal two-surface where lightrays focus as they enter the surface and expand on the other side once they have passed through the throat \cite{Hochberg:1998ha,McNutt}. Thus, the Weyl tensor is of algebraic type {\bf D} and the Ricci tensor is generally of algebraic type {\bf I} ($\Phi_{00} \neq 0$) relative to the alignment classification \cite{classb}. The conditions on the Cartan scalars for the AH and wormholes have been investigated in \cite{McNutt}.

\subsection{Dynamical solutions}

Recently exact solutions for evolving and collapsing spherically symmetric BD theory \cite{BT} have been found \cite{Pimentel}, in order to investigate the formation of black holes.
The	exact spherically symmetric solution interpreted
as an inhomogeneous dynamical scalar field collapsing cosmology, as studied in \cite{Husain}, is a special case (but written in the formally equivalent form of GR and a massless scalar field).
In particular, an event horizon can fail to form during the dynamical evolution
in the collapse phase in BD scalar field theory, leading to a naked singularity in the collapse endstate \cite{nakedZAT}.

There is a curvature singularity present at the AH,  $r=2M$, and another cosmological singularity. 
Since the metric is not static, the  AH is evolving with time. It is of interest to investigate the properties of the AH; e.g., whether the horizon is past or present, whether there exist trapping surfaces, etc. \cite{Pimentel,Husain}.
	
\subsection{Organization of the Paper}
Throughout this paper, we use geometric units in which $c=G=1$. We first establish the invariant characterization of the static spherically symmetric Brans Dicke models via the Cartan-Karlhede algorithm in Section 2, then we investigate the GR limit of such models and compare them to known static spherically symmetric models in GR in Section 3. We then investigate the properties of special geometrically defined surfaces in the models in Section 4, as well as investigate the invariant properties of the spacetime and how they compare to the Schwarzschild geometry in Section 5. In Section 6 we give an overview of the results of the analysis in the static spherically symmetric models, and in Section 7 we describe the applications of the same analysis to the dynamical Brans Dicke models. Finally, in Section 8, we briefly discuss the outcomes presented in this paper.

	\section{The Cartan-Karlhede Algorithm}
	An important problem in differential geometry is determining whether there exists a local coordinate transformation, $\Phi$, relating two spacetimes, $(M,g)$ and $({\bar M,\bar g})$; i.e.,
	
	\begin{equation}
		\Phi(\bar{g})=g.
	\end{equation}
	
	A method of determining if there exists such a transformation was proposed by Cartan, and later refined by Karlhede, by considering frames in which the metric components are constant, which we refer to as the Cartan-Karlhede algorithm, the steps of which we now list \cite{kramer}.
	
	\begin{enumerate}
		\item Set the order of differentiation $q$ to zero.
		\item Calculate the derivatives of the Riemann tensor up to the $q$th order.
		\item Find the canonical form of the Riemann tensor and its derivatives.
		\item Using Lorentz transformations of the frame, fix the frame as much as possible. Note the residual frame freedom (the group of allowed transformations, $I_q$ is the linear isotropy group) -- the number of allowed frame transformations that preserve the canonical forms. We will note dim$(I_q)$ as the dimension of the linear isotropy group.
		\item Find the number $t_q$ of independent functions the coordinates in the components of the Riemann tensor and its derivatives. The maximum for a spacetime is 4.
		\item If the isotropy group and number of independent functions are the same as in the previous step, let $p+1=q$ and stop; if they differ, increment $q$ by 1 and go back to step 2.
	\end{enumerate}
	
	We will denote $\mathcal{I}=\{R,R_{ab},R_{abcd},R_{abcd;e_1},...,R_{abcd;e_1...e_p} \}$, the set of components of the Riemann tensor and its derivatives up to the $p$th order. After computing the algorithm for a given spacetime, the elements of $\mathcal{I}$ are referred to as the Cartan scalars. We can then use these Cartan scalars, the successive isotropy groups, and the lists of the $t_q$ to compare two spacetimes. If all of these discrete properties match, then we can equate the corresponding Cartan scalars for each spacetime, and if there are solutions to those sets of equations, then we have shown there exists a coordinate transformation relating these two spacetimes. Thus the spacetimes are locally equivalent. An important point to note is that if any of the properties of the spacetimes in the algorithm, like the dimension of the isotropy groups, the $t_q$, and the numbers of Cartan scalars are different, then we immediately have that there is no equivalence between two such spacetimes. A detailed description of the Cartan-Karlhede algorithm (as well as the steps outlined above) is given in \cite{kramer}. To employ the Cartan Karlhede algorithm in this work, we will utilize the Newman-Penrose (NP) formalism. We also use the boost weight (bw) decomposition \cite{boostweight}, with the group of Lorentz transformations of NP null tetrads to analyze the properties of the curvature tensor.

	\section{Analysis of the Static Spherically Symmetric Models}
	We begin the analysis of the static spherically symmetric BD solution given in equations (\ref{bdss}) and (\ref{bdss_sf}). First we construct a null frame from the metric (\ref{bdss}), and employ the NP formalism and analyze the curvature tensors in the null frame.

	\subsection{Null Frame}\label{NF}
	We define the following orthonormal coframe $\{e_0,e_1,e_2,e_3 \}$ and null coframes $\{\ell_a,n_a,m_a,\bar{m}_a \}$, respectively, and derive the NP scalars in the null frame:

	\begin{equation}
	\begin{aligned}
		e_0 &= \left(1-\frac{2M}{r} \right)^{(m+1)/2}dt, \ e_1 = \left(1-\frac{2M}{r} \right)^{(n-1)/2}dr,\\
		e_2 &= r\left(1-\frac{2M}{r} \right)^{n/2} d\theta, \ e_3 = r\sin(\theta)\left(1-\frac{2M}{r} \right)^{n/2} d\varphi.
	\end{aligned}
	\end{equation}
	
	\begin{equation}
	\begin{aligned}
		\ell_a &= (e_1 - e_0)/\sqrt{2}, \ n_a = (e_1 + e_0)/\sqrt{2}, \\
		m_a &= (e_2 - ie_3)/\sqrt{2},\  \bar{m}_a = 	(e_2 + ie_3)/\sqrt{2}.
	\end{aligned}
	\end{equation}

	\noindent
	In this null frame, the non-zero NP scalars for the metric (\ref{bdss}) have the following coordinate expressions:
	\begin{equation}
		\Phi_{00} = \Phi_{22} = \left( 1-\frac{2M}{r}\right)^{-n-1} \frac{M\left(m M (n-2) + (m+n)r \right)}{2r^4},
	\end{equation}
	
	\begin{equation}
		\Phi_{11}{} = \left( 1-\frac{2M}{r}\right)^{-n-1} \frac{M\left( M(-m(m+5) + (m-3)n + n^2) + 2(m+n)r \right)}{4r^4},
	\end{equation}
	
	\begin{equation}
		\Psi_{2}{} =-\left( 1-\frac{2M}{r}\right)^{-n-1} \frac{M\left( M(3+m-n)(4+m-n) - 3(2+m-n)r   \right)}{6r^4},
	\end{equation}
	
	\begin{equation}
		\Lambda = -\left( 1-\frac{2M}{r}\right)^{-n-1} \frac{M^2 \left( m^2 + (n-1)n + m(n+1) \right)}{12r^4},
	\end{equation}

	\noindent
	The non-zero spin coefficients take the following coordinate expressions:
	
	\begin{equation}
	\begin{aligned}
		\rho &= \mu = -\left(1-\frac{2M}{r} \right)^{\frac{-1-n}{2}} \frac{(M(n-2) + r)}{\sqrt{2}r^2} ,\\
		\epsilon &= \gamma = \left(1-\frac{2M}{r} \right)^{\frac{-1-n}{2}} \frac{(m+1)M}{2\sqrt{2} r^2}
,\\
		\alpha &= -\beta = -\left(1-\frac{2M}{r} \right)^{\frac{-n}{2}}\frac{\cot(\theta)}{2\sqrt{2}r} .\\
	\end{aligned}
	\end{equation}

	\subsection{Zeroth Order}	
	At zeroth order, we compute the components of the Weyl and Ricci tensors in the invariant null frame presented, written as NP scalars. The Weyl tensor only contains a single component, $\Psi_2$, and the Ricci tensor contains three non-zero components, $\Phi_{00},\Phi_{11},\Phi_{22}$, two of which are equal, $\Phi_{00}=\Phi_{22}$, and we have the Ricci scalar,  $\mathcal{R}=24\Lambda$. All of these functions are only functions of a single coordinate, $r$; thus we have at zeroth order, $t_0=1$. The algebraically independent functions in the Riemann tensor at zeroth order are:
	
	\begin{equation}
		\begin{aligned}
			\Psi_2,\Phi_{00},\Phi_{11},\Lambda.
		\end{aligned}
	\end{equation}
	
	\noindent
	The isotropy group at zeroth order consists of spins, as boosts and null rotations will affect the form of the curvature tensors, which we wish to preserve at this step.

	\subsection{First Order} \label{CartanStaticFirst}
	There are no new functionally independent components at first order; $t_1=1$. The isotropy group at first order consists of only spins, as we have already used our boost and null rotation parameters to fix the frame at zeroth order, thus dim($H_1$)=$1$. We have the following relationship between Cartan scalars in the first derivative of the Weyl tensor:
	\begin{equation}
		C_{1432;1} = D\Psi_2 = -\Delta \Psi_2 = C_{1432;2}.
	\end{equation}

	\noindent
	None of the quantities in the curvature tensors are affected by spins due to the SO(3) symmetry inherent in the metric, thus we can set the spin parameter from our remaining isotropy freedom to vanish, completely fixing the frame, our null frame is now referred to as an invariant frame. The nonzero components of the Weyl tensor are listed in terms of the NP scalars and their frame derivatives:

	\begin{equation}
		\begin{aligned}
			C_{1212;1} &= -C_{1212;2} = -2 C_{1324;1} = 2C_{1342;2} =-2 C_{1423;1} = C_{3434;1}=-C_{3434;2}=2D\Psi_2,\\
			C_{1213;4} &= C_{1214;3} = C_{1223;4} = C_{1224;3} = -C_{1334;4} = C_{1434;3} =C_{2334;4} = -C_{2434;3} = 3\rho \Psi_2,
		\end{aligned}
	\end{equation}
	
	\noindent
	while the derivative of the traceless Ricci tensor, $S_{ab;c}$, contains the following non-zero scalars:
	
	\begin{equation}
		\begin{aligned}
			S_{11;1} = -S_{22;2} &=  8\Phi_{00}\epsilon - 2D\Phi_{00}, \\ \ S_{11;2} = -S_{22;1} &= 8\Phi_{00}\epsilon +2D\Phi_{00},\\
			S_{12;1} = S_{34;1} = -S_{12;2} = -S_{34;2} &= D \Phi_{11}, \\
			S_{13;4} = S_{14;3} = -S_{23;4} = - S_{24;3} &= 2\rho(\Phi_{00} -2\Phi_{11} ).
		\end{aligned}
	\end{equation}
	
		\noindent
	Using the Ricci and Bianchi identities, we can simplify all of the components of the Riemann tensor at zeroth, first and second order, eliminating almost all expressions involving frame derivatives. Additionally, we have multiple identities relating the curvature scalars and spin coefficients:
	
	\begin{equation}
		\begin{aligned}
			\Psi_2 &= -2\Lambda + \Phi_{00} + 4\epsilon \rho,\\
			D \Psi_2 + 2D\Lambda - D\Phi_{00} &= 4\Phi_{00}\epsilon + \rho(2\Phi_{11} + 3\Psi_2 - \Phi_{00}),\\
			D\Psi_2 - D\Lambda - D\Phi_{11} &= \rho(\Phi_{00} - 2\Phi_{11} + 3\Psi_2).
		\end{aligned}
	\end{equation}

	\noindent
	We complete the Cartan-Karlhede algorithm at first order, since no new functionally independent components appear, and the Cartan scalars appearing at second order will be the classifying functions. The complete set of algebraically independent Cartan scalars and extended Cartan scalars is then:
	
	\begin{equation}
		\Psi_2,\Phi_{11},\Phi_{00},\Lambda,\rho,\epsilon ~~\textrm{+ frame derivatives}. 
	\end{equation}

	\section{Limits to General Relativity}
	In the limit $|\omega|\to \infty$, and noting that $\frac{d\omega}{d\phi}=0$ trivially, the GR subcase of this family of solutions corresponds to $m=-n$, for which the BD parameter blows up. Applying this limit to the Cartan scalars of our invariant frame, we get the following:
	
	\begin{equation}
		\Phi_{00}{}=\Phi_{22}=-2\Phi_{11} =\frac{\mathcal{R}}{4}=-\left( 1-\frac{2M}{r}\right)^{-n-1}\frac{ M^2 (n-2) n}{2 r^4}	\end{equation}

	\begin{equation}
		\Psi_{2}{} = \left( 1-\frac{2M}{r}\right)^{-n-1}\frac{M(M (n-2) (2 n-3)+3 (n-1) r)}{3 r^4},
	\end{equation}
	
	\begin{equation}
		\begin{aligned}
			-\Delta\Psi_2 = D\Psi_2 &= 3\rho \Psi_2 + \frac23 \Phi_{00}(2\epsilon + \rho),\\
			-\Delta \Phi_{00} = D\Phi_{00} &= 4\Phi_{00}(-\epsilon + \rho),
		\end{aligned}
	\end{equation}
	\noindent
	where we have used the Bianchi identities to eliminate all occurrences of frame derivatives in the Cartan scalars. Comparing these Cartan scalars to the Cartan scalars derived for the Schwarzschild solution (below), we can see that the sets are all equivalent for the two cases $n=0,2$, indicating that the Schwarzschild geometry is possibly equivalent to one or both of these cases for the parameter $n$.

	Another way to examine this family of solutions is by looking at the relationships between the NP scalars; for example, we can look at the `perfect fluid' condition for the NP scalars,
	
	\begin{equation}
		\Phi_{00}\Phi_{22} = 4 \Phi_{11}^{\ 2}.
	\end{equation}

	\noindent
	In the GR limit, this condition is satisfied for all values of the parameter $n$, thus the solutions here correspond to perfect fluid solutions in GR, which is expected since there is an equivalent way of writing this spacetime as a scalar field solution in GR which has the form of a perfect fluid. Only in two cases ($n=0,2$) do we get vacuum solutions possibly related to the Schwarzschild geometry, while the rest of the family correspond to static scalar field solutions. Making the reparameterization $n\to 1-\gamma$ of this family of solutions, we recover the line element of the Janis-Newman-Winicour (JNW) solution \cite{Virbadhra97}:
	
	\begin{equation}
		ds^2 = -\left(1-\frac{b_{JNW}}{r} \right)^{\gamma}dt^2 + \left(1-\frac{b_{JNW}}{r}  \right)^{-\gamma}dr^2 + r^2\left(1-\frac{b_{JNW}}{r}  \right)^{1-\gamma}d\Omega^2,
	\end{equation}
	
	\noindent
	 where $d\Omega^2$ is the usual metric of the 2-sphere. For these geometries to be locally equivalent, we require the constant corresponding to the JNW solution satisfy $b_{JNW}=2M_{BD}$, where $M_{BD}$ is the constant in the BD solution. As a result of this relationship between the JNW solution and the BD solution in the GR limit, we can identify properties of the solutions in the JNW as also belonging to our BD solution in the appropriate GR limit. 

	The appearance of a Killing horizon at the surface $r=2M$ is only compatible with the parameter range for $m>-1$, which in the GR limit, corresponds to $n<1$. There are no Killing horizons for $n\ge1$ in the GR limit. 

	\section{Curvature Invariants in the n,m Parameter Space}
	The scalar polynomial invariants (SPI), $\{R,R^{ab}R_{ab},R^{abcd}R_{abcd},R^{abcd;e}R_{abcd;e},...\}$, in our coordinates, have the following expressions:

	\begin{equation}
		\mathcal{R} = \left( 1-\frac{2M}{r}\right)^{-n-1}\frac{4M^2(m^2+nm+n^2+m-1)}{2r^4},
	\end{equation}
	
	\begin{dmath}
	R_{ab}R^{ab}= \left( 1-\frac{2M}{r}\right)^{-2n-2}\frac{2 M^2}{r^8}	 \left(-2 M r (m+n) \left(m^2+m (7-2 n)-(n-3) n\right)+M^2
   \left(m^4+6 m^3+m^2 (2 (n-3) n+17)+2 m n ((n-4) n+7)+n^2 ((n-4) n+5)\right)+3 r^2
   (m+n)^2\right)
	\end{dmath}
	
	\begin{dmath}
		R_{abcd}R^{abcd} = \left( 1-\frac{2M}{r}\right)^{-2n-2}\frac{4 M^2}{r^8}\left(M^2 \left((3 m (m+2)+29) n^2-2 m (m (m+10)+17) n+m (m (m
   (m+10)+41)+56)+n^4-8 n^3-56 n+48\right)-4 M r \left(-2 m (m+2) n+m (m (m+8)+13)-n^3+6
   n^2-13 n+12\right)+6 r^2 (m (m+2)+(n-2) n+2)\right)
      \end{dmath}
	
	\begin{equation}
		R_{abcd;e}R^{abcd;e} \propto \left( 1-\frac{2M}{r}\right)^{-3n-3},
	\end{equation}
	
	\begin{equation}
		C_{abcd;e}C^{abcd;e} \propto \left( 1-\frac{2M}{r}\right)^{-3n-3},
	\end{equation}
	
	\noindent
	 where the differential invariants above are multiplied by long expressions involving $(r,m,n,M)$. It is clear from these expressions that these quantities are all finite at $r=2M$ when $n\le -1$. Additionally, all curvature invariants vanish on the surface $r=2M$ when $n< -1$ $\forall m$. We also notice that the parameter $m$ does not affect the occurrence of curvature singularities in any of these solutions. Thus we can conclude that there are curvature singularities for all solutions at $r=0$, and another curvature singularity for all solutions with $n>-1$, which appear in all of the above SPIs.

	We can further examine the Ricci scalar, and note that there is a subfamily of these static BD solutions with a vanishing Ricci scalar. We can solve the above expression for the parameters $(m,n)$ to determine when the Ricci scalar vanishes:
	
	\begin{equation}
		n=\frac{1}{2} \left(\pm\sqrt{-3 m^2-6 m+1}-m+1\right).
	\end{equation}
	
	\noindent
	These parameters trace out an ellipse in the $(m,n)$ parameter space, and in the GR limit, when $m=-n$, the only two vanishing Ricci scalar solutions are $n=0,2$, as expected.  
	
	\subsection{Example of the Identification of Geometrically Special Surfaces}
	As an example, we look at a specific BD solution, $m=-1,n=0$, and determine the expansion-free surface from the Cartan scalars. Additionally, the first order differential SPI in these cases identify the surface $r=2M$.
	
	\begin{equation}
	\begin{aligned}
		R &= 0,\\
		R_{ab}R^{ab} &= \frac{12M^2}{2r^6}, \\ 
	    R_{abcd}R^{abcd} &= \frac{24M^2}{r^6}, \\ \
	    R_{abcd;e}R^{abcd;e} &= \frac{360M^2\left(1-\frac{2M}{r}\right)}{r^{8}},\\
	    C_{abcd;e}C^{abcd;e} &= \frac{180M^2\left(1-\frac{2M}{r}\right)}{r^{8}}.
	\end{aligned}
	\end{equation}
	
	\noindent
	Notice that the differential curvature invariants above vanish on the surface $r=2M$. Moreover, the Cartan invariants $D\Psi_2, \rho$ also vanish on this surface, which means that all components of the first derivative of the Weyl tensor vanish on $r=2M$: 
	
	\begin{eqnarray}
		D\Psi_2 = \frac{3M \sqrt{1 -  \frac{2M}{r}}}{2 \sqrt{2} r^4}, \ \rho = - \frac{\sqrt{1 -  \frac{2M}{r}}}{\sqrt{2} r}.
	\end{eqnarray}

	\noindent
	In addition, the spin coefficients $\gamma=\epsilon=0$ since $m=-1$. We also notice that the number of algebraically independent components of the traceless Ricci tensor reduces, since for these parameters, $\Phi_{00} = \Phi_{22} = - \Phi_{11}$, and the frame derivative of these quantities also vanish on the $r=2M$ surface,
	\begin{equation}
		D\Phi_{00} = -D\Phi_{11} = \frac{3 \sqrt{2}M\, \sqrt{1 -\frac{2M}{r}}}{4 r^{4}},
	\end{equation}
	
	\noindent
	which tells us that the first derivative of the traceless Ricci tensor also completely vanishes on the $r=2M$ surface, yet the zeroth order tensors do not. All of the Cartan scalars belonging to the first order set of components thus detect the surface $r=2M$. This indicates that the Cartan scalars at first and second order detect the expansion-free surface at $r=2M$ in this geometry. Since all first-order Cartan scalars vanish on this surface, then so do the SPIs $C_{abcd;e}C^{abcd;e}, R_{ab;c}R^{ab;c},R_{abcd;e}R^{abcd;e}$, indicating that this surface is geometrically special, and of interest.

	\section{Schwarzschild Limit}
	
	The Schwarzschild limit of this class of solutions is when $m+n=0$ and the geometry becomes asymptotically flat and has a vanishing Ricci tensor and Ricci scalar. It is noted in \cite{campanelli93} that $m=n=0$ and $n=-m=2$ are two cases that are possibly locally related to a Schwarzschild geometry. To do a proper comparison of these two geometries, we will employ the Cartan-Karlhede algorithm again. For the Schwarzschild geometry, we use the following line element:
	
	\begin{equation}
		ds^2 = -\left(1-\frac{2M}{r} \right)dt^2 + \left(1-\frac{2M}{r} \right)^{-1}dr^2 + r^2(d\theta^2 + \sin(\theta)^2d\varphi^2).
	\end{equation}

	\subsection{Null Frame for Schwarzschild}
	We choose the initial frame to be aligned with the principal null directions of the Weyl tensor, partially fixing the frame:
	
	\begin{equation}
		\begin{aligned}
			\tilde{\ell} &= \frac{1}{\sqrt{2}}\left(\frac{\partial_t}{\sqrt{1-\frac{2M}{r}}} + \sqrt{1-\frac{2M}{r}}\partial_r \right), \\
			\tilde{n} &= \frac{1}{\sqrt{2}}\left(\frac{\partial_t}{\sqrt{1-\frac{2M}{r}}} - \sqrt{1-\frac{2M}{r}}\partial_r \right), \\
			\tilde{m} &= \frac{1}{\sqrt{2}}\left(\frac{\partial_{\theta}}{r} + i \frac{\partial_{\varphi}}{r\sin(\theta)} \right), \\
			\bar{\tilde{m}} &= \frac{1}{\sqrt{2}}\left(\frac{\partial_{\theta}}{r} + i \frac{\partial_{\varphi}}{r\sin(\theta)} \right).
		\end{aligned}
	\end{equation}

	\subsection{Cartan-Karlhede Algorithm}
	\noindent
	Since the Schwarzschild solution is a vacuum solution, we only need to consider the Weyl spinor components. At zeroth order, there is only a single component of the Weyl spinor. We also list the non-zero spin coefficients.
	
	\begin{equation}
		\tilde{C}_{1342}=-\frac{M}{r^3} = \Psi_2.
	\end{equation}

	\begin{equation}
		\begin{aligned}
			\rho&=\mu=-\frac{\sqrt{1-\frac{2M}{r}}}{\sqrt{2} r}, \\
			\epsilon &= \gamma =\frac{\sqrt{2}\, M}{4 r^2 \sqrt{1 -\frac{2M}{r}}}, \\
			\alpha &= -\beta = -\frac{\cos \! \left(\theta \right) \sqrt{2}}{4 \sin \! \left(\theta \right) r}.
		\end{aligned}
	\end{equation}

	\noindent
	At first order, there is also only one functionally independent component of the covariant derivative of the Weyl tensor, written as a function of the zeroth order components, and we write it in terms of coordinates, Cartan invariants, or NP scalars, respectively:
	\begin{equation}
		\tilde{C}_{1212;1} = \frac{6M\sqrt{r-2M}}{\sqrt{2}r^{9/2}}= \left(\frac{\Psi_2{}^{4}}{M}\right)^{1/3}\sqrt{\frac{1}{2} + \Psi_2{}^{1/3}M^{2/3}}  =D\Psi_2 = 3\rho\Psi_2
	\end{equation}

	\noindent
	Additionally, the Bianchi identities give the following relationship:
	
	\begin{equation}
		D\Psi_2 = 3 \rho \Psi_2.
	\end{equation}

	\noindent
	For the given frame, the Ricci identities produce another relationship between the curvature scalars and the spin coefficients:
	
	\begin{equation}
		\Psi_2 =  4 \epsilon \rho.
	\end{equation}

	\noindent
	Since there are no new independent functions of the coordinates, the algorithm stops here, and we continue to second order to determine the classifying functions.
	
	At second order, there are only two algebraically independent components of the second derivative of the Weyl spinor, and they can be written in terms of the zeroth order components and spin coefficients. We list the independent components in terms of coordinates, Cartan invariants, and NP scalars, respectively:

	\begin{equation}
		\begin{aligned}
			\tilde{C}_{1212;12} &=\frac{6M(3M-2r)}{r^6} = 18\Psi_2{}^2 + 12\frac{\Psi_2{}^{5/3}}{M^{2/3}}= 6\Psi_2 \left(-\Psi_2 +4\rho^2  \right)\\
			\tilde{C}_{1212;11} &= \frac{12M(r-2M)}{r^6}= -24\Psi_2{}^2 - 12\frac{\Psi_2{}^{5/3}}{M^{2/3}} = -24\Psi_2 \rho^2
		\end{aligned}
	\end{equation}

	Notably, all nonzero boost weight components of the second derivative of the Weyl tensor are directly proportional to $\tilde{C}_{1212;11}$, which vanishes on the surface $\rho=0 \ (r=2M)$, while the boost weight zero components, of which a number are proportional to $\tilde{C}_{1212;12}$, do not vanish on $r=2M$. Also notice that at first order, all components of the Weyl tensor are proportional to $3\rho\Psi_2$; this implies that the first order SPI, $\tilde{C}_{abcd;e}\tilde{C}^{abcd;e}$, vanishes on the horizon. The geometric horizon conjecture states \cite{COLEY2017131,GH} that the geometric horizons are surfaces for which the Riemann tensor (or its derivatives) is algebraically special compared to the external spacetime (i.e., the highest boost weight components of the tensor in the invariant null frame vanish).

\subsection{Local equivalence of the Brans-Dicke and Schwarzschild solutions}
From the derivation of the Cartan scalars for the BD solution and the Schwarzschild solutions, we can determine the local equivalence of the BD family in the GR limit ($m\to-n$) to the Schwarzschild solution. It is immediate that if $m\ne -n$, the set of Cartan scalars for the BD solution is distinct from the Schwarschild solution, thus these geometries are locally inequivalent. Now we want to establish under which circumstances the geometries are locally equivalent. In the case $n=m=0$, we have a one to one correspondence between the Cartan scalars of the BD solution and the Schwarzschild solution, determined up to second order (the maximal order required for these cases); thus the $m=-n=0$ solution is locally equivalent to the Schwarschild geometry.

\begin{table}[H]
\caption{Summary of the scalars appearing at zeroth order in the Riemann tensor for the static Brans Dicke solution. We can see that the cases $n=0,2$ eliminate all trace components of the Riemann tensor, indicating this may possibly be related to the Schwarzschild geometry, as we show using the Cartan-Karlhede algorithm. The $n\ne 0,2$ cases are inequivalent to the Schwarzschild geometry. The equivalence to Schwarzschild relies on identifying the equivalence between the constant parameters $M$ in either solution. We put a `hat' on the $n=2$ solution variables to differentiate the notation from $n=0$, and an `s' on the Schwarzschild coordinate, to differentiate its coordinate from the Brans Dicke family. }
\centering
\begin{tabular}{c|c|c|c|c}
\text{Scalar} & \text{Schwarzschild} & \text{Brans-Dicke $m=-n$} & \text{$m=-n=0$} & \text{$m=-n=2$}\\\hline
$\Psi_{2}{}$ & $- \frac{M}{ r_s^3}$ & $- \frac{M (2M (6 - 7 n + 2 n^2) + 6 (-1 + n) r)}{6 (1 -  \frac{2M}{r})^n (2M -  r) r^3}$ & $- \frac{M}{ r^3}$ & $ \frac{\hat{M}}{ (  \hat{r}-2\hat{M} )^3}$\\
$\Phi_{00}{}$ &$ 0 $&$ \frac{M^2 (-2 + n) n}{2(1 -  \frac{2M}{r})^n (2M -  r) r^3} $& 0 & 0\\
\end{tabular}
\end{table}

\vspace{2cm}
For the $n=0,2$ solutions, we can see that there exists a formal equivalence between the two solutions. However, they can only be formally equivalent if the parameter $M$ changes sign between the solutions. Consider $M,\hat{M}>0$, then the $n=0$ solution has a Schwarzschild horizon at $r=2M$, and a coordinate singularity at $r=0$. Conversely, the $n=2$ solution has no horizon, and as a coordinate singularity at $\hat{r}=2\hat{M}$ (note that the Kretschmann scalar is directly proportional to $\Psi_2{}^2$). But if we make the identification $M=-\hat{M}$, then the coordinate transformation $\hat{r} \to r - 2M$ will bring the $n=2$ solution into exactly the line element of the $n=0$ solution. These manifolds can be considered locally equivalent only if the constants appearing in their invariant classification are properly identified. This analysis extends to the full comparison of all parameters $(n,m)$ in the Cartan-Karlhede algorithm,

\subsection{Locally Equivalent Solutions}

In the GR limit ($m = -n$), we have established that two of the solutions in the family correspond to vacuum solutions, one of which is related to the typical Schwarzschild geometry, $n=0$, and the other related to the `negative mass' equivalent of the Schwarzschild geometry, $n=2$. Now the question arises: are the solutions with $n\ne 0,2$ describing different geometries? Noting that the solutions with $m=-n=0,2$ are formally related to the Schwarzschild geometry, and thus are related by the coordinate transformation $r \to 2\hat{M}-\hat{x}$, we can define a new coordinate $\hat{x}$ and transform the coordinate expressions in the Cartan scalars at zeroth order (in comparing two solutions, we relabel the $(m,n)$ parameters to $(q,p)$, and assign a hat to the variables for the second solution):

\begin{equation}
	\Phi_{00} = -\frac{M(M m(n-2) + (m+n)r)}{2r^4}\left(1-\frac{2M}{r}  \right)^{-n-1},
\end{equation}

\begin{equation}
	\widehat{\Phi}_{00} = -\frac{\hat{M}(\hat{M} p(-2-q)+(p+q)\hat{x}  )}{2\hat{x}^4}\left(1-\frac{2\hat{M}}{\hat{x}}  \right)^{p-3},
\end{equation}

\noindent
where we can see that the coordinate expressions for the components are related if we make the identification $p=-n+2$, $q=-m-2$. Thus, any two solutions with parameters $(m,n),(q,p)$ satisfying these relations are locally related by a coordinate transformation (the exact one above in fact). Let us augment this observation with a fully invariant analysis of all Cartan scalars.

We can show that the ratio of the two Cartan scalars $\Psi_2$ and $\mathcal{R}$ (when $\mathcal{R}\ne 0$) gives us a relationship for the coordinate $r$ in each member of the family, provided $m^2+m (n+1)+(n-1)
   n \ne 0$ (which corresponds directly to the subfamily of vanishing Ricci scalar solutions). This allows us to eliminate the coordinate $r$ for all Cartan scalars, and compare them in an invariant fashion. We will call this ratio of Cartan scalars $C$:

\begin{equation}
C=\frac{-M (m-n+3) (m-n+4)+3 (m-n+2)r}{12M \left(m^2+m (n+1)+(n-1)
   n\right)}.
\end{equation}

\noindent
 Now all of the Cartan scalars depend on the parameters $m,n,M$. To find equivalent solutions in this family, we need to find where transformations of these parameters leave the expressions invariant. If we transform the parameters $(n,m,M)$ according to $\tilde{n} \to -n +2$, $\tilde{m}\to -m-2$, $\tilde{M}\to -M$, then the Cartan scalars all keep their functional forms. Thus we conclude that there are local equivalences between members of the BD family of solutions, which are `negative mass' equivalent geometries. For example, the solutions with $m=-n$ and $n=3,-1$ are locally equivalent, $n=4,-2$ are locally equivalent, etc. Of course, this formal equivalence only holds if we carefully identify the constant $M$ between the solutions. If we only consider solutions with $M>0~ (<0)$, then by this same analysis, we can see that all members of this family of spacetimes with equivalent $M>0$ are distinct.

	\section{Special Surfaces in the Families of Solutions}
	For the appearance of expansion-free surfaces (possibly related to geometric horizons), we require the positive boost weight components of the Riemann tensor to vanish at all orders. Because of the static nature of these solutions, and the definition of the frame, there is a symmetry in the positive and negative boost weight components of the curvature tensors. In terms of the NP scalars, there are corresponding pairs of equivalences between all such quantities (see section \ref{NF}). We look for places where the non zero boost weight components vanish, these surfaces correspond to the vanishing of the extended Cartan invariant (the spin coefficient) $\rho$.
	
	\begin{equation}
		\rho = -\frac{1}{\sqrt{2} r^2}\left(1- \frac{2M}{r} \right)^{(-1-n)/2}  \left( M(n-2) + r \right),
	\end{equation}

	\noindent
	We can solve for the zeros of $\rho$ to get the following cases:
	
	\begin{table}[H]
	\caption{Roots of the extended Cartan invariant $\rho$. For all $-1 < n < \infty$, there exists curvature singularities in all of the SPIs given above, at the surface $r=2M$, with the exception of $n=0$. Solutions which are noted NS are ones which have naked singularities. The invariant $\rho$ is independent of the parameter $m$, so the appearance of an expansion-free surface (EFS) is defined for all $m$.}
	\centering
		\begin{tabular}{c|c|c|c}
		    $n$ & Roots of $\rho$ & Notes & Curv. Sing.\\ \hline
			$n\le -1$ & $r_1=2M$,$\ r_2=M(2-n)$ &  $r_2 >r_1$ & $r=0$ \\ 
			$ -1 < n < 0$ &  $r_{EFS} =M(2-n) $  & $r_{EFS} \in (2M, 3M) $ &  $r=0,2M$ \\ 
			$n =0$        & $r_{EFS}=2M$     & Double root at $r_{EFS}=2M$ &  $r=0$\\
			$0 < n \le 1$   & $r_{EFS} =M(2-n) $ & $r_{EFS} \in (0,2M)$, NS &  $r=0,2M$\\
			$1 < n < 2 $ & $r_{EFS}=M(2-n)$ & No EFS/KH, NS & $r=0,2M$\\
			$ n>2$ & No roots & No EFS/KH/Event Horizon, NS & $r=0,2M$
		\end{tabular}
	\end{table}

	\subsection{Inspection of the Cartan Scalars in the GR limit}
	Next we investigate the effect of the vanishing of $\rho$ in the GR limit on the components of the Riemann tensor at zeroth first and second order. First, let us simplify the components of the Riemann tensor using the Ricci and Bianchi identities. We get the following relationships between NP scalars and their derivatives:

	\begin{equation}
		\begin{aligned}
			\Psi_2=\frac23(\Phi_{00} + 4\epsilon \rho), \\
			D\Psi_2 = 3\rho \Psi_2 + \frac23\Phi_{00}(2\epsilon + \rho),\\
			D\Phi_{00} = 4\Phi_{00}(\rho - \epsilon), \\
			D\rho = \rho(2\epsilon+\rho) + \Phi_{00}.
		\end{aligned}
	\end{equation}

\noindent
We can now simplify the components of the Riemann tensor, using these identities. We have omitted all quantities which are proportional to any of these quantities. At zeroth order, we have only components that are bw $-2,0,2$. We list the algebraically independent components (the bw $-2$ components are equal to the bw $+2$ components):
\begin{equation}
	\begin{aligned}
		\textrm{bw 2: }R_{1341}&=\Phi_{00},\\
		\textrm{bw 0: } R_{1221}&=-2(\Phi_{00}- 4\epsilon \rho),\\
		R_{1342}&=\Phi_{00}+4\epsilon \rho, \\
		R_{3443}&=\frac23(2\Phi_{00}-3\Psi_2).\\
	\end{aligned}
\end{equation}

\noindent
At first order, we have components of bw $\pm 1,~\pm 3$. Because of the symmetry in the frame components, the bw $\pm 1$ terms are equivalent and similarly, the bw $\pm 3$ terms are equivalent. Let us list the algebraically independent bw components:
\begin{equation}
	\begin{aligned}
		 \textrm{bw 3: } R_{1341;1}&=4\Phi_{00}(2\epsilon - \rho) ,\\
		 \textrm{bw 1: }R_{1221;1}&=2\rho(\Phi_{00}+3\Psi_2),\\
		 R_{1231;4}&=-3\Psi_2\rho,\\
		 R_{3443;1}&=-2(3\rho\Psi_2 - \Phi_{00}(\rho - 2\epsilon)),\\
		 R_{1341;2}&=4\Phi_{00}\rho. \\
	\end{aligned}
\end{equation}

At second order, we have components of bw 4,2,0,-2,-4. Again, due to the symmetry, the set of bw $\pm 4,\pm2$ components are equal. There is only one unique component of bw $\pm 4$, eleven unique bw $\pm 2$ components, and nine bw zero components. We list here a few of the Cartan scalars at second order, and omit the rest:

\begin{equation}
	\begin{aligned}
		\textrm{bw 4: } R_{1341;11}&=\frac43\Phi_{00}\left(5\Phi_{00} - 3(\Psi_2 + 24\epsilon^2 - 16\epsilon \rho + 5\rho^2) \right),\\
		\textrm{bw 2: } R_{1221;11}&=2\Phi_{00}(2\Phi_{00}+3\Psi_2) - 4(7\Phi_{00}+6\Psi_2)\rho^2, \\
		\textrm{bw 0: } R_{1232;14}&=-2(\Phi_{00} + 6\Psi_2)\rho^2.
	\end{aligned}
\end{equation}

\section{Dynamical Brans Dicke Model}
Let us study the exact dynamical and generalized solution given by \cite{Pimentel} of CL type \cite{campanelli93}:
	
	\begin{dmath}\label{bdss2}
		ds^2 = - (at+b)^{1-k}\left(1-\frac{2M}{r}\right)^{m+1}dt^2 + (at+b)^{1-k}\left(1-\frac{2M}{r} \right)^{n-1}dr^2 + r^2(at+b)^{1-k}\left(1-\frac{2M}{r}\right)^{n}(d\theta^2 + \sin(\theta)^2d\phi^2),
	\end{dmath}
	
	\begin{equation}
		\varphi = \varphi_0 (at+b)^{k}\left(1-\frac{2M}{r} \right)^{-(m+n)/2},
	\end{equation}
	
	\noindent
	where $m,n,M,a,b$ are constants, and $k$ is related to the BD \cite{BT} parameter $\omega$ by
	
	\begin{equation}\label{CL_twoparameters}
		\omega = -2\frac{m^2 + n^2 + nm + m - n}{(m+n)^2},\ k=\sqrt{\frac{3}{3+2\omega}}=\sqrt{-\frac{3(m+n)^2}{(m-n) (m-n+4)}}.
	\end{equation}
	
	\noindent
		This solution is a conformal rescaling of the $\omega\rightarrow\infty$ limit of the CL solution (\ref{bdss}), but for a restricted parameterization. The constants $(m,n,k)$ are related to the constants $ (M,K,J)$ of \cite{Pimentel} via
	
	\begin{equation}
		n=1-2K-M, \ m=M-1,\ k=J, 
	\end{equation}
	
	\noindent
	and, further, are connected to the BD parameter $\omega$ through
	
	\begin{eqnarray}
		&n = \frac{1}{2} \left(-\sqrt{\frac{1}{2 \omega +3}}-\sqrt{3}+2\right), \\ &m= \frac{1}{2} \left(-\sqrt{\frac{1}{2 \omega
   +3}}+\sqrt{3}-2\right).
	\end{eqnarray}

	\noindent
	We solve the field equations with the $(m,n)$ parameters and show that there is only a single parameter solution for the BD field equations in terms of the parameter $\omega$, or by transforming to the variables used by CL we get a solution in terms of the parameter $n$ (or $m$).	Writing our solution in terms of the parameter $n$, we get the subsets of solutions given in Table \ref{bdparams}.

	\begin{table}[H]
	\centering
	\caption{Dynamical Brans Dicke solution parameters written in terms of the CL parameter $n$. These correspond to three solutions, the first two are a reparameterization of Pimentel's solution \cite{Pimentel}, and the third is the limit to Husain's solution \cite{Husain}. }
	\begin{tabular}{c|c|c}\label{bdparams}
	k & m & $\omega$  \\ \hline
	 $\sqrt{3}(-2+\sqrt{3} +2n)$ & $-2 + \sqrt{3} + n $ & $\frac{1}{2} \left(\frac{1}{\left(2 n+\sqrt{3}-2\right)^2}-3\right)$ \\
	 $\sqrt{3}(2+\sqrt{3} -2n)$ &  $-2 - \sqrt{3} +n$ & $\frac{1}{2} \left(\frac{1}{\left(-2 n+\sqrt{3}+2\right)^2}-3\right)$\\
	$0$ & $-n$ &  $-\infty$
	\end{tabular}\label{BDdynamicParameters}
	\end{table}

	The metric (\ref{bdss2}) is a solution for the parameters $(M,a,b,\omega)$. The valid range for the coordinates are $2M < r <\infty$, and $-b/a \le t < \infty$. The $r=2M$ locus is a timelike curvature singularity, and the point $t=-b/a$ is a spacelike curvature singularity. The solution with $M=0$ corresponds to a spatially homogeneous and isotropic cosmology for all other parameter values.

	\subsection{GR limit of the four-parameter dynamical Brans Dicke solution}
	
	For the four-parameter solution to the BD field equations, in the limit $\omega \rightarrow \infty$, which results in $k=0$, we obtain the following metric and scalar field solutions:
	
	\begin{equation}
		ds^2 =  (at+b)\left(-\left(1-\frac{2M}{r}\right)^{-n+1}dt^2 + \left(1-\frac{2M}{r} \right)^{n-1} + r^2\left(1-\frac{2M}{r}\right)^{n}(d\theta^2 + \sin(\theta)^2d\varphi^2)\right),
	\end{equation}
	
	\noindent
	where the BD scalar field is now constant everywhere:
	\begin{equation}
		\phi = \phi_0.
	\end{equation}

	\noindent
 This metric is conformally related to a static solution of \cite{campanelli93} with conformal factor $(at+b)$ and is of the general form of the models studied by Husain \cite{Husain} (but in the equivalent form of a massless scalar field in GR). The locus of points at $r=2M$  is a spacelike curvature singularity, and so the `horizon' is shrunk to a point as in the static case as discussed above. There is also a timelike (cosmological) singularity at $t=-b/a$. The coordinate ranges and the values of the constants are (for the metric to be Lorentzian): $-b/a \le t \le \infty$ and $ 2M\le r\le \infty$ ($M>0$). We note that for $a\ne 0$, there is a coordinate transformation that eliminates the constant $b$. But we are  also interested in the metrics for which $a=0$, so we will not consider those redefinitions of $t$. We note also that when $b=0$, there are two ranges for $t$, namely $ 0 \le t \le \infty$ for $a>0$, and $ -\infty \le t \le 0$ for $a<0$, corresponding to white and black hole-like solutions, respectively (and simply reflect the choice of the arrow of time). $M=0$ corresponds to a spatially homogeneous and isotropic cosmology.

\subsection{Null Frame}

We start by constructing the null frame by choosing the principal null directions of the Weyl tensor as the vectors $\ell$ and $n$:

\begin{equation}
	\begin{aligned}
		\ell &= \left( \frac{\partial_t}{A} + \frac{\partial_r}{B}\right)/\sqrt{2},\\
		n &= \left( \frac{\partial_t}{A} - \frac{\partial_r}{B}\right)/\sqrt{2},\\
		m &= \left( \frac{\partial_{\theta}}{R} + i\frac{\partial_{\phi}}{R\sin(\theta)}\right)/\sqrt{2},\\
		\bar{m} &= \left( \frac{\partial_{\theta}}{R} - i\frac{\partial_{\phi}}{R\sin(\theta)}\right)/\sqrt{2},\\
	\end{aligned}
\end{equation}

\noindent
where the functions $A,B,R$ have the following forms

\begin{equation}
	\begin{aligned}
		A(t,r)&=(at+b)^{(1-k)/2}\left(1-\frac{2M}{r}\right)^{(m+1)/2},\\
		B(t,r)&=(at+b)^{(1-k)/2}\left(1-\frac{2M}{r}\right)^{(n-1)/2},\\
		R(t,r)&=r(at+b)^{(1-k)/2}\left(1-\frac{2M}{r}\right)^{n/2}.\\
	\end{aligned}
\end{equation}

\noindent
The spin coefficients in this frame have the following coordinate expressions (where we use the abbreviations $F\equiv (1-2M/r)$ and $ ~T\equiv (at + b)$ for the sake of brevity):

\begin{equation}
\rho=\frac{T^{\frac{k -3}{2}} F^{\frac{1}{2} (-m-n-1)}}{2 \sqrt{2} r^2} \left(-2 T F^{m/2} (M (n-2)+r) + a  (k -1) r^2
   F^{n/2}\right),
\end{equation}

\begin{equation}
	\mu = \frac{T^{\frac{k -3}{2}} F^{\frac{1}{2} (-m-n-1)} }{2 \sqrt{2} r^2}\left(-2 T F^{m/2} (M (n-2)+r)-a 
   (k -1) r^2 F^{n/2}\right),
\end{equation}

\begin{equation}
	\epsilon = \frac{T^{\frac{k -3}{2}} F^{\frac{1}{2} (-m-n-1)}}{4 \sqrt{2} r^2} \left(2 (m+1) M T F^{m/2}-a 
   (k -1) r^2 F^{n/2}\right),
\end{equation}

\begin{equation}
	\gamma = \frac{T^{\frac{k -3}{2}} F^{\frac{1}{2} (-m-n-1)}}{4 \sqrt{2} r^2} \left(2 (m+1) M T F^{m/2}+a 
   (k -1) r^2 F^{n/2}\right),
\end{equation}

\begin{equation}
	\alpha =-\beta= -\frac{F^{-n/2} \cot (\theta ) T^{\frac{k -1}{2}}}{2 \sqrt{2} r}.
\end{equation}

\noindent
The NP curvature scalars have the following coordinate expressions:

\begin{equation}
	\Lambda = \frac{T^{k -3}}{48 F r^4} \left(3 a ^2 \left(k ^2-1\right) r^4 F^{-m}-4 M^2 T^2 F^{-n}
   \left(m^2+m (n+1)+(n-1) n\right)\right),
\end{equation}

\begin{dmath}
	\Phi_{00} = \frac{T^{k-3} F^{-m-n-1}}{8 r^4} \left(a^2 (k-3) (k-1) r^4 F^n-4 a (k-1) (m+1) M r^2 T
   F^{\frac{m+n}{2}}+4 M T^2 F^m (m M (n-2)+r (m+n))\right),
\end{dmath}

\begin{dmath}
\Phi_{22}=\frac{T^{k-3} F^{-m-n-1} }{8 r^4}\left(a^2 (k-3) (k-1) r^4 F^n+4 a (k-1) (m+1) M r^2 T
   F^{\frac{m+n}{2}}+4 M T^2 F^m (m M (n-2)+r (m+n))\right),
\end{dmath}

\begin{dmath}
	\Phi_{11}=\frac{T^{k-3}}{16 F r^4} \left(a^2 (k-3) (k-1) r^4 F^{-m}-4 M T^2 F^{-n} \left(M \left((m-3) n-m
   (m+5)+n^2\right)+2 r (m+n)\right)\right),
\end{dmath}

\begin{equation}
	\Psi_2=\frac{M F^{-n-1} T^{k-1} }{6 r^4}(M (m-n+3) (m-n+4)-3 r (m-n+2)),
\end{equation}

\noindent
where we have left the constants $k,m$ in symbolic form. Any NP scalar or NP spin coefficient not listed here vanishes in this frame.

\subsection{Cartan Karlhede Algorithm}
We employ the Cartan Karlhede algorithm here for the dynamical BD models, using the above NP scalars and null frame.
\subsubsection{Zeroth Order}
At zeroth order, we have only the quantities $\{ \Psi_2,\Phi_{00},\Phi_{11},\Phi_{22}, \Lambda\}$ appearing in the components of the Riemann tensor, two of which can be taken as the independent scalars, with the remaining scalars depending on these. We take $\Psi_2$ and $\Phi_{11}$ to be the functionally independent Cartan scalars at the zeroth order. Since we cannot explicitly solve for $t$ and $r$ for these expressions, we will use these scalars as an implicit function basis for the Cartan scalars in this frame. Additionally, these functions can be shown to be functionally independent since the wedge product of their exterior derivatives is nonvanishing:

\begin{equation}
	d\Psi_2 \wedge d\Phi_{11} \ne 0, \implies \nexists \ G(\Psi_2,\Phi_{11})=0.
\end{equation}

\noindent
where $G$ is an arbitrary function of the $\Psi_2,\Phi_{11}$. We can choose our boost parameter to set $\Phi_{00}=\Phi_{22}$, the remaining isotropy freedom at this order is only spins (dim$(H_0)=1$).

\subsubsection{First Order}
\noindent
Relative to the NP scalar and frame derivatives, the form of the Weyl and trace-free Ricci tensors is the same as in section \ref{CartanStaticFirst}. No new functionally independent quantities appear at first order, and the algorithm terminates. We must continue to second order to determine the classifying functions for the spacetimes. The frame, with the corresponding boost applied, is now an invariant frame.

\subsection{Scalar Polynomial Invariants for the Dynamical Models}
Inspecting the Ricci and Kretschmann scalars for the dynamical BD solution, we get the following expressions:

\begin{equation}\label{dynRicciScalar}
	\mathcal{R}=\frac{2 \left(3 n^2+3 \left(\sqrt{3}-2\right) n-3 \sqrt{3}+5\right) F^{-n-\sqrt{3}-1} T^{2
   \sqrt{3} (n-1)} \left(3 a^2 F^2 r^4-F^{\sqrt{3}} M^2 T^2\right)}{r^4},
\end{equation}

\begin{equation}
	R_{abcd}R^{abcd}\propto \frac{F^{-2(1+n)}T^{4\sqrt{3}(n-1)}}{r^{10}} \mathcal{G}(t,r),
\end{equation}

\noindent
where $\mathcal{G}(t,r)$ is some long expression of the coordinates and parameters, and we have used the first solution for $(k,m)$ in Table [\ref{BDdynamicParameters}]. From this expression, there exists curvature singularities at $r=2M$ for $n>-\sqrt{3}-1$. There is also a cosmological singularity at $t=-b/a$ for $n<1$. Moreover, there is also a set of dynamical BD models for which there is a vanishing Ricci scalar. We can solve $\mathcal{R}=0$ for $n$ to get:

\begin{equation}
	n=\frac13(3-2\sqrt{3}),~\frac13(3-\sqrt{3}),
\end{equation}

\noindent
thus there are two solutions for which the Ricci scalar vanishes. There are no solutions for which the Ricci tensor also vanishes, so we do not find the Schwarzschild solutions in the GR limit of this dynamical BD family of solutions. The only GR limit of these vanishing Ricci scalar solutions is to the scalar field collapse solution found in \cite{Husain}.

\section{Discussion}
In the analysis of the Brans-Dicke static spherically symmetric solutions, we have found that the two-parameter family of solutions contains a number of distinct families of solutions, identified by their canonical forms in the invariant frame formalism. A subset of the parameter range in the GR limit is related to other members by a formal coordinate transformation, though are different in the signs of their arbitrary constants. These families correspond to Minkowski $(M=0)$, Schwarzschild $(m=-n=0)$, `negative mass' Schwarzschild\footnotemark $(m=-n=2)$, perfect fluid solutions equivalent to GR with a scalar field source $(m=-n\ne 0,2)$, a family of Brans-Dicke solutions with a spin symmetric Ricci tensor $(m\ne-n, ~ \Phi_{00} \ne a \Phi_{11}, ~a~\rm{const})$, and another small subfamily of these with vanishing Ricci scalar BD solutions. Interestingly, the only solutions for which the Ricci tensor completely vanishes in the family of BD static solutions correspond only to the two formally equivalent Schwarzschild cases, which holds for all $m,n$. This agrees with \cite{Hawk} in comparing the GR limits to stationary vacuum solutions in GR. The results of this identification of the subfamilies of the static spherically symmetric BD solutions are listed in Table \ref{families}.

	\begin{table}[H]
	\centering
	\caption{Subfamilies of solutions contained within the static and dynamical spherically symmetric Brans Dicke solutions. BD here indicates a solution is a solution to the BD field equations, and GR indicates the solution is the GR limit of the BD solution. The tag -S indicates a solution is from the static BD family, and the tag -D indicates the solution is from the dynamical BD family. VRS stands for ``Vanishing Ricci Scalar".}
	\begin{tabular}{c|c}\label{families}
	Subfamily & Identifying Properties  \\ \hline
	 (Conformal) Minkowski (BD-S,D) & $M=0$\\
	 VRS (GR-S) &  $\mathcal{R}=0$, $R_{ab}\ne0$ \\
	Schwarzschild (see footnote) (GR-S) &$m=-n=0,2$, $\mathcal{R}=0,~R_{ab}=0$ \\
	JNW (GR-S) & $m=-n \ne 0,2$, $\mathcal{R}\ne0,~R_{ab}\ne0$\\
	Spin Symmetric (BD-S) & $m\ne -n$, $\Phi_{00} \ne a \Phi_{11}, ~a~\textrm{const}$\\
	Collapsing Scalar Field (GR-D) & $m=-n$, $k$ given in last row of Table \ref{BDdynamicParameters}\\
	Dynamical BD Vacuum (BD-D) & $m\ne -n$, $k$ given in first two rows of Table \ref{BDdynamicParameters}
	\end{tabular}
	\end{table}

\vspace{1cm}
 We find that the appearance of the expansion-free surface ($\rho=0$) in the GR limit depends on the parameter $n$, where we require $n\le -1$ for the spin coefficient  $\rho$ to vanish, where we get two $\rho=0$ surfaces here. The two surfaces identified by zero $\rho$ exist for a small range of the parameter $n$, and for $n>2$ no horizons or Killing horizons are detected by any invariants, thus resulting in the occurrence of naked singularities in these spacetimes.

In the case of the generalized dynamical models, we see similar properties to the static cases in the appearance of expansion-free surface, which correspond to apparent horizons in the GR case \cite{Husain}. These models also contain naked singularities for a certain parameter range of $n$, and additionally contain cosmological singularities at different times, depending on the parameters $a,b$. We also determine that the only GR limit of the dynamical BD spacetimes corresponds to the scalar field collapse solution in \cite{Husain}. We list the subfamilies of the dynamical BD solution in Table \ref{families}.

 It is clear that the Cartan invariants allow a way of determing properties of interest in a spacetime, here we establish their usefulness for identifying naked singularities, black holes, and wormholes. In addition, the Cartan invariants offer a way to determine the local equivalence of spacetimes, something which SPIs are unable to do.

\footnotetext{The $n=2$ solution is not necessarily a Schwarzschild solution, depending on the identification of the constant parameter; rather it is a horizonless `negative mass' Schwarzschild solution if we only consider $M>0$.}

\section*{Acknowledgments}
AAC is supported by the Natural Sciences and Engineering Research Council of Canada. DD is supported by an AARMS fellowship. We also thank Valerio Faraoni for helpful comments on Brans Dicke and JNW solutions.

\section*{Declarations}
AAC is supported by the Natural Sciences and Engineering Research Council of Canada. DD is supported by an AARMS fellowship.


\begin{thebibliography}{20}

\bibitem{Dey1} 
  K.~Mosani, D.~Dey and P.~S.~Joshi,
  ``Strong curvature naked singularities in spherically symmetric perfect fluid collapse,'' Phys.\ Rev.\ D {\bf 101}, no. 4, 044052 (2020).

\bibitem{Dey2} D.Dey, P.S.Joshi, K.Mosani and V.Vertogradov, ``Causal structure of singularity in non-spherical gravitational collapse,'' Eur. Phys. J. C \textbf{82}, no.5, 431 (2022).
 
 
\bibitem{BT} C. Brans and  R. H. Dicke, “Mach’s Principle and a Relativistic Theory of Gravitation”, Phys. Rev. 124, 925 (1961).

\bibitem{Hochberg:1998ha} D. Hochberg and M. Visser. Geometric structure of the generic static traversable wormhole throat. Phys. Rev. D, 56:4745–4755 (1997). arXiv:gr-qc/9704082 [gr-qc].
	
\bibitem{COLEY2017131} A. Coley and D. McNutt. Identification of black hole horizons using scalar curvature invariants. Classical and Quantum Gravity, 35(2):025013 (2017).

\bibitem{classb} R. Milson, A. Coley, V. Pravda, and A. Pravdova, Int. J. Geom. Meth. Mod. Phys., 2(01):41–61, 2005. arXiv:0401010 [gr-qc].

\bibitem{GH} A. A. Coley, D. D. McNutt, and A. A. Shoom. Geometric horizons. Physics Letters B, 771:131–135, 2017; D. Brooks, P. C. Chavy-Waddy, A. A. Coley, A. Forget, D. Gregoris, M. A. H. MacCallum, and D. D. McNutt, Cartan invariants and event horizon detection. Gen. Relativ. Gravit., 50:37 (2018) [arXiv:1709.03362 [gr-qc]].

\bibitem{kramer} H. Stephani, D Kramer, M. MacCallum, C. Hoenselaers, and E. Herlt. Exact Solutions of Einstein’s Field Equations. Cambridge University Press (2009).

\bibitem{McNutt} D. D. McNutt, W. Julius, M. Gorban, B. Mattingly, P. Brown and G. Cleaver, Geometric surfaces: An invariant characterization of spherically symmetric black hole horizons and wormhole throats, Phys. Rev. D 103, 124024 (2021).

\bibitem{Kozak} A. Kozak and A. Wojnar, Eur. Phys. J. vC81, 492 (2021).

\bibitem{mcpage} D. D. McNutt and D. N. Page. Scalar polynomial curvature invariant vanishing on the event horizon of any black hole metric conformal to a static spherical metric. Phys. Rev.D, 95(8):084044 (2017). arXiv:1704.02461 [gr-qc].

\bibitem{Paiva} F. M. Paiva and C. Romero, On the limits of Brans-Dicke spacetimes: a coordinate free approach, [arXiv:gr-qc/9304031].

\bibitem{Letelier} P. S. Letelier and A. Wang, Phys. Rev D 48, 631 (1993)

\bibitem{Brans62} C. H. Brans , Phys. Rev. 125,2194 (1962).

\bibitem{Hawk}  S.W. Hawking, Commun. Math. Phys. 25, 167 (1972).

\bibitem{val}  V. Faraoni, F. Hammad and S.D. Belknap-Keet, Phys. Rev. D 94, 104019 (2016); V. Faraoni, F. Hammad, A. M. Cardini, and T. Gobeil, Phys Rev D 97, 084033 (2018).

\bibitem{campanelli93} M. Campanelli and C.O. Lousto, Are black holes in brans-dicke theory precisely the same as in general relativity? Int. Journal of Mod. Phys. D 02, 451, (1993).

\bibitem{Bhadra2005} Bhadra, A., Sarkar, K. On static spherically symmetric solutions of the vacuum Brans-Dicke theory. Gen Relativ Gravit 37, 2189–2199 (2005). 

\bibitem{Agnese95} A. G. Agnese and  M. La Camera, Wormholes in the brans-dicke theory of gravitation. Physical review. D, 51(4):2011–2013 (1995).

\bibitem{Bron} K. A. Bronnikov, Acta Phys. Polon. B 4, 251 (1973)

\bibitem{morris88} M. S. Morris and K. S. Thorne, Wormholes in spacetime and their use for interstellar travel:
A tool for teaching general relativity, Am. J. Phys., 56, 395 (1988).

\bibitem{Visser:1995cc} 
M. Visser, Lorentzian Wormholes: From Einstein to Hawking. Springer-Verlag (1995).

\bibitem{Pimentel} L. O. Pimentel, Mod. Phys. Lett A, vol 12, 1865 (1997)

\bibitem{Husain} V. Husain, E. A. Martinez, D. N\'u\~nez, Exact solution for scalar field collapse, Phys Rev D.50.3783 (1994) [arXiv:gr-qc/9402021]. 


\bibitem{boostweight} A. Coley, R. Milson, V. Pravda and A. Pravdova, Class. Quant. Grav. 21, L35 (2004) [gr-qc/0401008]; A. Coley, R. Milson, V. Pravda and A. Pravdova, Class. Quant. Grav. 21, 5519 (2004) [gr-qc/0410070].


\bibitem{nakedZAT}  A. H. Ziaie, K. Atazadeh and Y. Tavakoli, Class. Quant. Grav. v27 075016 \& 209801 (2010)

\bibitem{Virbadhra97} K. S. Virbhadra, Janis Newman Winicour and Wyman Solutions are the Same, International Journal of Modern Physics A (1997). 


\end{thebibliography}
\end{document}